\documentclass[letterpaper,pra,aps,floatfix,superscriptaddress,onecolumn,longbibliography,notitlepage
]{quantumarticle}
\pdfoutput=1
\usepackage[utf8]{inputenc}
\usepackage[english]{babel}
\usepackage[T1]{fontenc}
\usepackage{amsmath,amsthm,amsfonts,amssymb,xcolor}
\usepackage{graphicx}
\usepackage{microtype}
\usepackage{braket}
\usepackage[numbers,sort&compress,square]{natbib}

\usepackage{subcaption}

\makeatletter
\def\l@subsubsection#1#2{}
\makeatother

\usepackage[colorlinks=true,citecolor=blue,linkcolor=blue]{hyperref}
\usepackage[capitalize,nameinlink]{cleveref}

\theoremstyle{definition}

\newcommand{\rmdist}{\mathrm{dist}}

\newcommand{\expec}{\mathbb{E}}

\newcommand{\be}{\begin{equation}}
\newcommand{\ee}{\end{equation}}

\begin{document}

\title{A Classical Algorithm Which Also Beats $\frac{1}{2}+\frac{2}{\pi}\frac{1}{\sqrt{D}}$ For High Girth MAX-CUT}

\author{Matthew B.~Hastings}                                                                                                                                                                                                                    \affiliation{Station Q, Microsoft Quantum, Santa Barbara, CA 93106-6105, USA}                                          \affiliation{Microsoft Quantum and Microsoft Research, Redmond, WA 98052, USA}                                                                                                                                                                                                                                                              

\begin{abstract}
We give a simple classical algorithm which provably achieves the performance in the title.  The algorithm is a simple modification of the Gaussian wave process\cite{lyons2017factors}.
\end{abstract}
\maketitle

MAX-CUT is a well-studied optimization problem.  On high girth graphs with degree $D$, there are a variety of classical algorithms which achieve approximation ratio $\frac{1}{2}+\frac{2}{\pi}\frac{1}{\sqrt{D}}$ up to subleading corrections.  Examples include the Gaussian wave process
\cite{lyons2017factors} and also \cite{thompson2021explicit}.

 There is also a classical algorithm due to Montanari and collaborators\cite{montanari2019optimization,alaoui2020optimization,new} which, assuming one conjecture, beats this performance and indeed achieves
arbitrarily close to the optimal performance.  This conjecture is generally believed to hold by the spin-glass community.  However, this ratio 
 $\frac{1}{2}+\frac{2}{\pi}\frac{1}{\sqrt{D}}$ is the best approximation
 ratio that we know for a classical algorithm that can be proven without additional assumptions; see \cite{thompson2021explicit}.

Recently, it was pointed out\cite{p11maxcut} that $p=11$ QAOA attains approximation ratio
$\frac{1}{2}+C\frac{1}{\sqrt{D}}$ on high girth graphs for some $C>2/\pi$.
The point of the present short note is to show that a very simple modification
of the Gaussian wave process can also achieve some other $C>2/\pi$.

We make no attempt to optimize the performance.
However, since numerical effort has been expended to optimize the performance of QAOA, it seems that at least as much effort should be expended to optimize the performance of local classical algorithms, by performing numerical integrations to determine their performance and optimize parameters.
Indeed, the algorithm of \cite{hirvonen2014large} is a single round local algorithm that was obtained using computer optimization.

\section{The Problem and Previous Algorithms}
We consider a graph with fixed degree $D$.  We assume that the girth of the graph is large.  We consider the problem MAX-CUT: maximizing the fraction of edges which are cut by a bipartition of the vertices.

We define a bipartition of the vertices by assigning a sign, $\pm 1$, to each vertex.  We call that sign the ``part" of that vertex and denote it $p_u$ for vertex $u$.  An edge is cut if it connects two vertices with opposite sign. 

If we assign the signs independently from the uniform distribution, then each edge is cut with probability $1/2$.

The best one can achieve in general for large $D$ is to find a cut where a fraction $1/2+C/\sqrt{D}+o(1/\sqrt{D})$ of edges are cut.  Since in this short note we only care about the performance at large $D$ and large girth (doing anything else would require something more complicated than this simple calculation), we care only about achieving a large value of $C$.

The algorithm of \cite{new} is shown (under a widely-believed conjecture) to achieve arbitrarily close to the Parisi\cite{parisi1979toward,dembo2017extremal} value of $C$, $C=0.763\ldots\ $  Note $2/\pi=0.6366\ldots$

\section{The Algorithm}
We now give the algorithm.  In \cref{gwave} we give a simple algorithm that achieves arbitrarily close to $C=2/\pi$.  This algorithm is essentially the same as that in \cite{lyons2017factors}.
In \cref{aoms}, we show how to improve beyond $C=2/\pi$.

\subsection{Arbitrarily Close to $2/\pi$}
\label{gwave}
There is a real variable for each vertex.  Initialize the real variables by choosing them to be independent Gaussian, all with zero mean and variance $1$.
Label vertices $u,v,\ldots$ and label the corresponding real variables $x_u,x_v,\ldots$.

Then, in parallel update each $x_u$ to some new $x'_u$.  A simple update rule which achieves $C>0$ is
\be
\label{update1}
x'_u = x_u + a \sum_{v, \, \rmdist(u,v)=1} x_v,
\ee
where $a$ is a real scalar and the notation $\rmdist(u,v)=1$ indicates the sum over all $v$ which neighbor $u$.  Here $\rmdist(u,v)$ is the graph metric.

Finally, round each variable $x'_u$ by assigning each vertex $u$ to the part $p_u\sigma(x'_u)$, where $\sigma(\cdot)$ denotes the sign function.

Choosing $a$ to be negative and proportional to $1/\sqrt{D}$ gives a positive value of $C$.

In a slightly more complicated algorithm, everything is the same except we replace \cref{update1} with
\be
\label{update2}
x'_u = \sum_{k=0}^K a_k \sum_{v,\, \rmdist(u,v)=k} x_v,
\ee
for some given integer $k$ and some sequence of reals $a_k$.
Note that for $a_0=1, a_1=a$, \cref{update2} reduces to \cref{update1}.

We claim that by choosing the $a_k$ appropriately, we may take
\be
\label{expectratio}
\frac{\expec[x'_u x'_v]}{\expec[(x'_u)^2]}=-\frac{2}{\sqrt{D}} \cdot(1-o(1)),
\ee
for neighboring $u,v$,
where the $o(1)$ is a quantity tending to zero as $D,K\rightarrow \infty$.
One may see this by considering the spectrum of the adjacency matrix of an infinite tree. However, a simple explicit choice is $a_k=(-1/\sqrt{D})^k$, up to a multiplicative factor for normalization. In this case,
$\expec[(x'_u)^2]=\sum_{k=0}^K D^{-k} N_k=K\cdot(1-o(1))$, where $N_k= D^k\cdot(1-o(1))$ is the number of vertices at distance $k$.
At the same time, for neighboring $u,v$, \begin{eqnarray}
&&\expec[x'_u x'_v]\\ \nonumber&=&\sum_{w,\, \rmdist(u,w)= \rmdist(v,w)+1 \leq K} (-1/\sqrt{D})^{\rmdist(u,w)+\rmdist(v,w)}
+\sum_{w,\, \rmdist(v,w)= \rmdist(u,w)+1 \leq K} (-1/\sqrt{D})^{\rmdist(u,w)+\rmdist(v,w)}\\ \nonumber&=&-(2/\sqrt{D})(K-1)\cdot(1-o(1))=(-2/\sqrt{D}) K \cdot(1-o(1)).\end{eqnarray}

From \cref{expectratio} it follows that\cite{goemans1995improved}
\be
\expec[\sigma(x'_u)\sigma(x'_v)]=-\frac{2}{\pi} \frac{2}{\sqrt{D}}\cdot(1-o(1)).
\ee

Finally, the expected cut fraction equals $1/2-(1/2)\expec[\sigma(x'_u)\sigma(x'_v)]=
1/2+(2/\pi) \frac{1}{\sqrt{D}} \cdot(1-o(1)).$

\subsection{And One More Step}
\label{aoms}
We now modify the algorithm further.
The initialization is as before, and we use \cref{update2} to define $x'_u$.
However we then choose an $\epsilon$-fraction of vertices uniformly at random, for some $\epsilon>0$, and ``mark" those vertices.
For each unmarked vertex $u$, we let $p_u=\sigma(x'_u)$.  For each marked vertex $u$, we compute the majority of $\sigma(x'_v)$ over $v$ which neighbor that vertex (if there is a tie, choose the majority arbitrarily) and then return $-1$ times that majority as the part.

In words: round all $x'_u$ to signs.  Then, on an $\epsilon$-fraction of vertices, perform a greedy update, flipping the part if it improves the cut.

Let us analyze the performance of this algorithm.

{\it Neither vertex marked---}
For any neighboring pair $u,v$, there is a $(1-\epsilon)^2$ probability that both vertices are unmarked, in which case $\expec[p_u p_v]=-\frac{2}{\pi} \frac{2}{\sqrt{D}}\cdot(1-o(1))$.

{\it One vertex marked---}
There is a probability $2\epsilon(1-\epsilon)$ that one vertex is marked and the other is unmarked.  Suppose $u$ is marked and $v$ is unmarked.
Then $\expec[p_u p_v]$ is \emph{at least as negative} as if both were unmarked, i.e.,  $\leq -\frac{2}{\pi} \frac{2}{\sqrt{D}}\cdot(1-o(1))$, because the flip of the marked vertex $u$ is done greedily and so it cannot increase $\sum_{w,\, \rmdist(u,w)=1} \expec[p_u \sigma(x'_w)]$, and since $v$ is unmarked, $p_v=\sigma(x'_v)$.

Further, we claim that $\expec[p_u p_v]=-c' \frac{2}{\sqrt{D}}\cdot(1-o(1))$, for some $c$ strictly greater than $2/\pi$.  To see this, consider the distribution of $x'_w$ for $w$ which neighbor $u$.  
These $x'_w$ are correlated Gaussian variables.
Normalize so that $\expec[(x'_u)^2]=1$.
The correlation matrix $\expec[x'_{w_1} x'_{w_2}]$ is a $(D-1)$-by-$(D-1)$ matrix, with diagonal entries equal to $1$ and all off-diagonal entries equal to
each other.  Off-diagonal entries are equal to $\Theta(1/D)$; in general, for arbitrary vertices $y,z$ the correlation is $(-1)^{\rmdist(y,z)} \Theta(1/\sqrt{D})^{\rmdist(y,z)}$.
We can generate such a correlation matrix by the following process: generate
independent 
identically distributed random Gaussian with zero mean and variance $1-o(1)$ for each $w$ neighboring $u$, and generate
 a
random Gaussian $g$ with zero mean and variance $\Theta(1/D)$.  Then, add $g$ to the variable on each vertex, giving $x'_w$ for the given vertex $w$.

So, conditioned on $g$, the signs $\sigma(x'_w)$ for $w$ which neighbor $u$ are independent random variables.  They are identically distributed, with a bias which is $\Theta(g)$, i.e., they have a probability $1/2+\Theta(g)$ of being the same as $\sigma(x'_u)$.  With probability at least $0.99$,   $g$ is bounded by some constant multiple of $1/\sqrt{D}$, in which case the sum of biases over neighbors is comparable 
to the root-mean-square variance in the sum of the signs, i.e., both are $\Theta(\sqrt{D})$.  Hence,
there is a probability $\Omega(1)$ that we flip $p_u$ in the last round of the algorithm, i.e., that $p_u = - \sigma(x'_u)$, and that further, when we do flip, the expectation value of the absolute value of sum of signs is $\Theta(\sqrt{D})$.
Hence, the improvement in cut, summed over neighbors, by flipping $p_u$ is
$\Theta(\sqrt{D})$.

So, the
claim follows that $\expec[p_u p_v]=-c' \frac{2}{\sqrt{D}}\cdot(1-o(1))$, for some $c'$ strictly greater than $2/\pi$.

{\it Both Vertices Marked---}
With probability $\epsilon^2$, both $u,v$ are marked.
We claim that $\expec[p_u p_v]\leq O(1/\sqrt{D})$.  Likely the expectation value is negative and the expectation value could be estimated by some numerical integration, but we are doing the minimum amount of work here.

Indeed, suppose $\expec[p_u p_v]= \omega(1/\sqrt{D})$.  Then, we could define a different algorithm which took $\epsilon=1$ (i.e., it marks all vertices) and
which replaced our given choice of $a_k$ in \cref{update2} by $a_k (-1)^k$, and then we claim $\expec[p_u p_v]=-\omega(1/\sqrt{D})$, asymptotically better than the optimum!

To see this, imagine flipping the sign of $x_w$ for $\rmdist(u,w)$ odd.  This has no effect on the distribution of $x_w$.  Then, the change in $a_k$ also flips the sign of $x'_w$ for $\rmdist(u,w)$ odd, and the final greedy update flips the sign of $p_w$ for $\rmdist(u,w)$ \emph{even}.

{\it Performance of Algorithm---}
We have
\be
\expec[p_u p_v]\leq -(1-\epsilon)^2 \frac{2}{\pi} \frac{2}{\sqrt{D}}\cdot(1-o(1))
-2\epsilon(1-\epsilon)c' \frac{2}{\sqrt{D}}\cdot(1-o(1))+\epsilon^2 O(1/\sqrt{D}).
\ee
Up to subleading terms, this is a quadratic function of $\epsilon$, with all coefficients proportional to $1/\sqrt{D}$ and the linear term is negative.
Hence, for some constant $\epsilon>0$, the algorithm achieves
$\expec[p_u p_v]\leq -C \frac{2}{\sqrt{D}} \cdot(1-o(1))$ for some $C$ strictly greater than $2/\pi$.

\subsection{Related Algorithms}
This algorithm is similar to \cite{Hastings_2019,bapat2020approximate}.  There, a variable was initialized between $[-1,+1]$ on each vertex.  One then repeats the following two-step process:
update each variable with a linear function of its value and the sum of the values on its neighbors, and then replace the value of the variable $x_u$ on each vertex $u$ with $\tanh(\beta x_u)$ for some $\beta$.  
After some number of rounds, finally round each variable by replacing it with its sign.  The linear functions and $\beta$ could be taken to be functions of the round.

Suppose we replace the initialization with Gaussian variables instead of uniformly distributed ones.
Keeping $\beta$ small for the early rounds, this dynamics remains in the linear regime, and by iterating it we reproduce 
\cref{update2} up to small nonlinearities.  Taking a large value of $\beta$ for the next round implements the rounding to a value close to $\pm 1$.  The ``one more step" can be done if we allow the injection of randomness: add random noise to each variable so that there is a small chance that each variable changes from close to $\pm 1$ to close to $0$.  Then, implement one more linear transform and round again.

Perhaps even without modification, this algorithm achieves $C>2/\pi$.  Other natural variants of the algorithm considered here include, for example, rounding the $x'_u$, and then following with a threshold algorithm which flips the sign if sufficiently many (compared to $\sqrt{D}$) of the neighbors have the same part, or adding some low order nonlinearities before rounding.  Numerical calculation can help optimize performance.

{\it Acknowledgments---} I thank K. Marwaha for pointing out \cite{new} and other comments.

\bibliography{refs}
\end{document}